\documentclass[sigconf]{acmart}

\usepackage{graphicx}

\AtBeginDocument{%
  }

\copyrightyear{2024}
\acmYear{2024}
\setcopyright{acmlicensed}
\acmConference[Conference acronym 'XX]{Make sure to enter the correct conference title from your rights confirmation emai}{June 03--05, 2018}{Woodstock, NY}
\acmISBN{978-1-4503-XXXX-X/18/06}
\acmDOI{XXXXXXX.XXXXXXX}
\begin{document}

\title{Mamba Retriever: Utilizing Mamba for Effective and Efficient Dense Retrieval}


\author{Hanqi Zhang}
\affiliation{%
  \department{Gaoling School of Artificial Intelligence}
  \institution{Renmin University of China}
  \city{Beijing}
  \country{China}
}
\email{zhanghanqi@ruc.edu.cn}

\author{Chong Chen}
\affiliation{%
  \institution{Huawei Cloud}
  \city{Beijing}
  \country{China}
}
\email{chenchong55@huawei.com}

\author{Lang Mei}
\affiliation{%
  \department{Gaoling School of Artificial Intelligence}
  \institution{Renmin University of China}
  \city{Beijing}
  \country{China}
}
\email{meilang2013@ruc.edu.cn}

\author{Qi Liu}
\affiliation{%
  \department{Gaoling School of Artificial Intelligence}
  \institution{Renmin University of China}
  \city{Beijing}
  \country{China}
}
\email{liuqi\_67@ruc.edu.cn}

\author{Jiaxin Mao}
\authornote{Corresponding author.}
\affiliation{%
  \department{Gaoling School of Artificial Intelligence}
  \institution{Renmin University of China}
  \city{Beijing}
  \country{China}
}
\email{maojiaxin@gmail.com}

\renewcommand{\shortauthors}{Hanqi Zhang, Chong Chen, Lang Mei, Qi Liu, and Jiaxin Mao}

\begin{abstract}
In the information retrieval (IR) area, dense retrieval (DR) models use deep learning techniques to encode queries and passages into embedding space to compute their semantic relations. It is important for DR models to balance both efficiency and effectiveness. Pre-trained language models (PLMs), especially Transformer-based PLMs, have been proven to be effective encoders of DR models. However, the self-attention component in Transformer-based PLM results in a computational complexity that grows quadratically with sequence length, and thus exhibits a slow inference speed for long-text retrieval. Some recently proposed non-Transformer PLMs, especially the Mamba architecture PLMs, have demonstrated not only comparable effectiveness to Transformer-based PLMs on generative language tasks but also better efficiency due to linear time scaling in sequence length. This paper implements the Mamba Retriever to explore whether Mamba can serve as an effective and efficient encoder of DR model for IR tasks. We fine-tune the Mamba Retriever on the classic short-text MS MARCO passage ranking dataset and the long-text LoCoV0 dataset. Experimental results show that (1) on the MS MARCO passage ranking dataset and BEIR, the Mamba Retriever achieves comparable or better effectiveness compared to Transformer-based retrieval models, and the effectiveness grows with the size of the Mamba model; (2) on the long-text LoCoV0 dataset, the Mamba Retriever can extend to longer text length than its pre-trained length after fine-tuning on retrieval task, and it has comparable or better effectiveness compared to other long-text retrieval models; (3) the Mamba Retriever has superior inference speed for long-text retrieval. In conclusion, Mamba Retriever is both effective and efficient, making it a practical model, especially for long-text retrieval.

\end{abstract}

\ccsdesc[500]{Information systems~Retrieval models and ranking}

\keywords{information retrieval, pretrained language models, state space model}


\maketitle

\section{Introduction}
Information retrieval (IR) aims to retrieve information objects that are relevant to users' queries from a large-scale collection.
Dense retrieval (DR) models \cite{karpukhin2020dense} are proposed to assess the relevance between query and passage by encoding them into embeddings and calculating the similarity in the embedding space. 
Then, using Approximate Nearest Neighbor Search algorithms \cite{shrivastava2014asymmetric} on embeddings, we can retrieve the top-k relevant passages to the query.

DR model needs to balance both effectiveness and efficiency.
Regarding effectiveness, DR models focus on improving retrieval performance, which is influenced by factors including model's inherent ability in semantic understanding and summarization. 
Regarding efficiency, this paper focuses on the passage inference time instead of query, which has greater potential for improvement.

Pre-trained Language Models (PLMs) have demonstrated their effectiveness on downstream tasks, thanks to the sufficient world knowledge and semantic knowledge they gain during pre-training.
Especially, the Transformer-based PLMs \cite{vaswani2017attention,kenton2019bert,touvron2023llama} represent the advantage of capturing long-range dependencies by the self-attention mechanism and allowing parallel training.
Many studies \cite{karpukhin2020dense,DBLP:conf/emnlp/Ni0LDAMZLHCY22,muennighoff2022sgpt,ma2023fine} have proposed to adopt Transformer-based PLMs as the encoders of DR models and have observed their effectiveness.

However, despite the effectiveness of Transformer-based DR models, the efficiency is limited by their inherent model architectures. 
In detail, the self-attention component in Transformer-based PLMs results in a computational complexity that grows quadratically with sequence length. 
Thus, for retrieval tasks with long passages, such as legal case retrieval task \cite{li2023lecardv2}, Transformer-based DR models exhibit slow inference speeds.



Some non-Transformer PLMs \cite{peng2023rwkv, fu2024monarch, gu2023mamba} are proposed to improve efficiency without sacrificing effectiveness.
In particular, Mamba-based PLMs present comparable performance to Transformer-based PLMs on generative language tasks and achieve linear time scaling in sequence length based on a selective state space model mechanism.
Recently, Xu proposed the RankMamba model \cite{xu2024rankmamba}, which leverages Mamba for re-ranking tasks. Our work differs in two key aspects: (1) The proposed Mamba Retriever is a bi-encoder model designed for effectiveness and efficiency in first-stage retrieval tasks. (2) We extend the investigation of Mamba-based models to long-text retrieval tasks.

In this paper, we propose to explore whether Mamba can serve as an effective and efficient encoder of DR model for IR tasks. We answer the following research questions:


\textit{RQ1: Does Mamba retrieval model have comparable effectiveness on classic retrieval compared to Transformer retrieval models?}

\textit{RQ2: Does Mamba retrieval model have comparable effectiveness on long-text retrieval compared to existing long-text retrieval models?}

\textit{RQ3: How does the inference efficiency of the Mamba retrieval model compare to existing retrieval models across various text lengths?}

To address these research questions, we implement the Mamba Retriever, a bi-encoder retrieval model based on Mamba. 
We fine-tune it on MS MARCO passage ranking dataset \cite{DBLP:conf/nips/NguyenRSGTMD16} for classic short-text retrieval and on LoCoV0 dataset \cite{saad2024benchmarking} for long-text retrieval.

We make the following contributions: 
(1) We implement the Mamba Retriever in order to achieve both effectiveness and efficiency. 
(2) We explore the retrieval effectiveness of Mamba Retriever at different model sizes. We show that, on the MS MARCO passage ranking dataset and BEIR 
\cite{thakur2021beir} datasets, Mamba Retriever has comparable or better effectiveness compared to Transformer retrievers, and the effectiveness also grows with the size of the Mamba model.
(3) Besides, we focus on the effectiveness of long-text retrieval. We show that, on the long-text LoCoV0 dataset, Mamba Retriever can extend to longer text length than its pre-trained length after fine-tuning on retrieval task, and it has comparable or better effectiveness compared to other long-text retrieval models.
(4) We explore the passage inference efficiency of Mamba Retriever at different passage lengths. We show that Mamba Retriever has superior inference speed with linear time scaling for long-text retrieval. 

In conclusion, Mamba Retriever is both effective and efficient, making it practical for IR, especially for long-text IR tasks.
More details are available at \href{https://github.com/41924076/MambaRetriever}{https://github.com/41924076/MambaRetriever}.

\section{Related Work}
\textbf{Pre-trained Language Models (PLMs). }
Through pre-training, language models can achieve higher performance when transferred to specific tasks. Transformer \cite{vaswani2017attention}, based on self-attention mechanism, is a mainstream architecture of PLMs, including encoder-only\cite{kenton2019bert,liu2019roberta} model and decoder-only\cite{zhang2022opt,biderman2023pythia} model. To address the quadratic time scaling of Transformer architecture, some architectures broadly regarded as state space models have been proposed\cite{peng2023rwkv,gu2021efficiently}, especially high-performing models like sub-quadratic architecture M2-BERT\cite{fu2024monarch}, linear architecture Mamba\cite{gu2023mamba} and Mamba-2\cite{dao2024transformers}. 

\textbf{Dense Retrieval Models. }
Transformer PLMs have been proven effective for dense retrieval. Initially, encoder-only models are adopted for retrieval tasks due to the bi-directional attention mechanisms \cite{karpukhin2020dense,DBLP:conf/emnlp/Ni0LDAMZLHCY22}. Later, decoder-only models are adopted for retrieval tasks due to their effectiveness on larger model size \cite{muennighoff2022sgpt,ma2023fine}.

\textbf{Long-text Dense Retrieval Models. }
In long-text retrieval, early works use chunking strategies \cite{dai2019deeper} due to the small context window. In order to help the model better understand complete and coherent semantics, some studies explore Transformer-based long-text retrieval models\cite{gunther2023jina,ma2023fine,zhu2024longembed}.

Faced with the quadratic time scaling of Transformer-based long-text retrieval models, the sub-quadratic M2-BERT model\cite{fu2024monarch} has been utilized for long-text retrieval tasks\cite{saad2024benchmarking}.


\section{Methodology}

\subsection{Mamba Retriever}\label{sec:Mamba Retriever}
\textbf{Task Definition. }
In text retrieval tasks, given a query \( q \)  and a large-scale passage set \( \{p_1, p_2, \ldots, p_n\} \) , the retrieval model aims to find top-\( k \) passages that are most relevant to \( q \). 


\textbf{Overview of Mamba Retriever. }
To calculate the relevance between a query \( q \) and a passage \( p \), Mamba Retriever uses a bi-encoder architecture. Bi-encoder means that the model represents \( q \) and \( p \) as dense vector embeddings \( E_q \) and \( E_p \) respectively, and relevance between the query \( q \) and passage \( p \) can be computed by the cosine similarity between their dense representations.
\begin{equation}
    sim(q,p) = \frac{E_p \cdot E_q}{\|E_p\| \|E_q\|}
\end{equation}

Specifically, to generate embedding \( E \), we use auto-regressive language model Mamba as base model \( M \). 
We input a token sequence \( t_1,t_2,\ldots,t_L \) which has a sequence length of \(L\) and a token <EOS> at the end of the sequence to the model \( M \), and extract the output of <EOS> at the last hidden layer in \( M \) as embedding \( E \):
\begin{equation}
    E = M(t_1,t_2,\ldots,t_L,\text{<EOS>})[-1]
\end{equation}

\textbf{Mamba and SSM. }
The base model Mamba, as we previously denoted as \( M \), is a model constructed by stacking multiple Mamba blocks. The core component of the Mamba block is selective state space model which is based on state space model (SSM)\cite{gu2021efficiently}.

A SSM maps 1-dimensional input
sequence $x(t) \in \mathbb{R}$ with time step $t$ to output sequence $y(t) \in \mathbb{R}$ through latent state $h(t) \in \mathbb{R}^N$:
\begin{align}
    h'(t) &= A h(t) + B x(t)  & y(t) &= C h(t)
\end{align}

where $A \in \mathbb{R}^{N \times N}$, $B \in \mathbb{R}^{N \times 1}$, $C \in \mathbb{R}^{1 \times N}$.
Using step size \(\Delta\), the continuous form above can be changed to a discrete form:
\begin{align}
    h_t &= \overline{A} h_{t-1} + \overline{B} x_t & y_t &= C h_t
    \label{eq:ht}
\end{align}
where $\overline{A} = \exp(\Delta A)$ and $\overline{B} = (\Delta A)^{-1} (\exp(\Delta A)-I) \cdot \Delta B$ is one of the discretization methods.
The above 1-dimensional SSM can be extended to independent \(d\) dimensions.

Based on the SSM, Mamba introduces a selection mechanism and corresponding hardware-aware parallel algorithm. 
The selection mechanism is making $\Delta, B, C$ dependent on the current token. It
allows the model to selectively forget or remember information along the dimension of sequence length.

\textbf{Training Objective. }
To train the Mamba Retriever, we employ the InfoNCE loss, which is most commonly used. It forces \( q \) and \( d \) that are semantically similar to be closer in embedding space:
\begin{equation}\label{eq:dpr_obj}
\centering
    \mathcal{L} =
    -\log \frac{e^{\text{sim}(q, d^+)/ \tau}}{e^{ \text{sim}(q, d^+)/ \tau} + \sum\limits_{d_i^-\in D^-} e^{\text{sim}(q, d_i^-)/ \tau}}
\end{equation}
where $d+$ is the relevant passage of the query, $D^-$ is a set of irrelevant passages of the query, and $\tau$ is the temperature coefficient.

\subsection{Base Model Comparison}\label{sec:mamba_vs_decoder}

In Section~\ref{sec:Mamba Retriever}, we let Mamba be the base model \( M \) of retriever.
As comparison, the base model $M$ is changed from Mamba to other frequently-used base models, including the Transformer encoder-only models and decoder-only models. 
In this section, we analyze the differences between using Mamba and these models as \( M \).

\textbf{Mamba and Decoder-only vs Encoder-only. }
Mamba and Transformer decoder-only models share similarities that distinguish them from Transformer encoder-only models.
In terms of data, Mamba and decoder-only models are pre-trained on more data than most encoder-only models. In particular, Mamba and Pythia \cite{biderman2023pythia} are pre-trained on the same data. 
In terms of architecture, Mamba and decoder-only models have causal characteristics,
which is not as suitable as encoder-only models with bi-directional attention for comprehension tasks like retrieval. Mamba can be reconstructed to be bi-directional, but this would lead to a decrease in efficiency. 

\textbf{Mamba vs Decoder-only. }
Intuitively, Transformer decoder-only model can capture long-term dependencies by self-attention mechanism, while Mamba may be limited by the maximum amount of information that can be compressed in latent states.

However, some works\cite{ali2024hidden,cirone2024theoretical,merrill2024illusion,dao2024transformers} analyze that Mamba has some mechanism similar to or even surpassing Transformer: Mamba has implicit attention mechanism with good expressiveness; 
if each SSM is regarded as one head in multi-head self-attention mechanism, then Mamba has more heads than the Transformer; 
the softmax in self-attention can cause problems, 
such as over-smoothing, whereas Mamba does not use softmax and thus may better capture subtle differences between different tokens.

In addition, Mamba has an additional explicit process of summarizing previous information using the latent states. 
When calculating a token at position \(t\), decoder-only model uses the attention mechanism to access keys and values of all previous tokens. However, Mamba uses the SSM mechanism to incorporate the information of all previous tokens in a fixed-size latent state \( h_{t-1} \) which serves as a summerization, and then calculates the output of position \(t\) using the \( h_{t-1} \) according to formula~\ref{eq:ht}. 
We can have an assumption that Mamba may acquire good summarizing abilities during pre-training, which is beneficial for retrieval where the goal is to summarize query or passage information into an embedding.





\section{Experiments} 
\subsection{Experiments on MS MARCO}\label{sec:ms}
\textbf{Dataset.}
We fine-tune and evaluate on MS MARCO passage ranking dataset \cite{DBLP:conf/nips/NguyenRSGTMD16}, a classic short-text retrieval dataset with the value of average passage char length less than 400. 
MRR@10 and Recall@1k are common metrics. 

Additionally, we conduct zero-shot evaluation on 13 BEIR datasets \cite{thakur2021beir} including ArguAna, Climate-FEVER, DBPedia, FEVER, FiQA, HotpotQA, NFCorpus, NQ, Quora, SCIDOCS, SciFact, TREC-COVID, Tóuche-2020. Average nDCG@10 is a common metric. 


\textbf{Implementation Details.} We initialize our model from a pre-trained Mamba checkpoint and train it on 4 $\times$ 64G V100 GPUs. 
Since efficiency is important for retrieval tasks, we conduct experiments on the model sizes of 130M, 370M, and 790M. 
We add a new special token <EOS> to the end of the text as a dense representation. We normalize the dense representation with L2 normalization, and fix the temperature coefficient at 0.01. Because the purpose of the experiment is not to use various tricks to achieve optimal performance, we use a batch size of 2 on each GPU. By sharing negative passages between GPUs and batches, each query totally has 63 negative passages mined by BM25. 
We use a well-performing learning rate 1e-5 and train until convergence. We train and evaluate on MS MARCO with query maximum length of \{16, 32\} respectively and passage maximum length of 128. 
We conduct zero-shot evaluations on BEIR with both query and passage maximum length of 64.

For baseline, the base model $M$ mentioned in Section~\ref{sec:Mamba Retriever} is changed from Mamba to Transformer models, including encoder-only model BERT \cite{kenton2019bert} and RoBERTa \cite{liu2019roberta}, and decoder-only model OPT \cite{zhang2022opt} and Pythia \cite{biderman2023pythia}. The training setup is basically the same as Mamba, with two key differences: the learning rate is set to the most effective value for each model respectively; BERT and RoBERTa use the last hidden state of prefix <CLS> token as the text embedding. 

\begin{table}[t]
\centering
\caption{The effectiveness of Mamba Retriever on MS MARCO and BEIR compared to Transformer Retrievers. Arch denotes architecture, R@1k is Recall@1k, En is Encoder-only, De is Decoder-only, and Ma is Mamba.}
\label{tab:ms}
\resizebox{\linewidth}{!}{
\begin{tabular}{l|c|c|cc|c}
    \toprule
    \textbf{Base Model} &  \textbf{Size}&  \textbf{Arch.}&  \multicolumn{2}{c|}{\bf MS MARCO} & \textbf{BEIR} \\
     &  &  &  \textbf{MRR@10} & \textbf{R@1k} & \textbf{nDCG@10} \\
    \midrule
    BERT-base     & 110M &En.& \textbf{32.8} & 95.2 &36.45\\
    RoBERTa-base     & 125M &En.& 31.3& 95.0&37.02 \\
    OPT      & 125M &De.& 31.1&  94.9&36.83\\
    Pythia     & 160M &De.& 25.4& 91.0 &31.47\\
    Mamba    & 130M &Ma.& 32.3& \textbf{96.7}&\textbf{40.54} \\
    \midrule
    BERT-large     & 330M &En.& 33.9& 95.7 &37.31\\
    RoBERTa-large     & 355M &En.&33.9 & 96.1&38.37 \\
    OPT      & 350M &De.&31.1 & 94.4 &35.89\\
    Pythia     & 410M &De.&31.9 & 96.6 &38.62\\
    Mamba    & 370M &Ma.& \textbf{35.2}& \textbf{97.7} &\textbf{43.52}\\
    \midrule
    Pythia     & 1B &De.& 33.8& 97.4 &43.11\\
    Mamba    & 790M &Ma.& \textbf{36.3}& \textbf{98.3} &\textbf{44.72}\\
    \midrule
    OPT     & 1.3B &De.&35.8 & 98.1 &42.87\\
    \bottomrule
\end{tabular}
}
\end{table}

\textbf{Results.} 
Table~\ref{tab:ms} shows the effectiveness on MS MARCO passage ranking and BEIR datasets. Briefly, Mamba shows comparable or better performance to Transformer and improves with model size.

When compared with Transformer decoder-only models, Mamba shows better performance on all model sizes, especially Pythia which is pre-trained on the same data. 
Transformer encoder-only models have higher scores than decoder-only, which proves that bi-directional self-attention is beneficial to retrieval.
When compared to the bi-directional Transformer encoder-only model, although Mamba is uni-directional, Mamba still has similar or better performance. 
This suggests that Mamba has stronger text comprehension and summarization ability, possibly due to its advantages discussed in Section~\ref{sec:mamba_vs_decoder}, such as Mamba's implicit attention mechanism and explicit summarization ability.
Additionally, Mamba's retrieval performance grows with the model size expanding from 100M to 790M.


\subsection{Experiments on LoCoV0}
\textbf{Dataset.}
Our models are trained and evaluated on LoCoV0 dataset. The LoCoV0 has 5 long-text retrieval datasets: SummScreenFD, Gov Report, QMSUM, QASPER Title to Full Text, and QASPER Abstract to Full Text. The average char lengths of passages are \{30792, 55280, 58129, 22315, 22315\}. Average nDCG@10 is a common metric. The recently released LoCoV1 consists of more datasets, but was not released 
at the time of conducting our experiments.


\textbf{Implementation Details.} 
We initialize our model from a pre-trained Mamba checkpoint and train on 8 $\times$ 64G V100 GPUs. We use a maximum length of 2k or 8k tokens. Training and evaluation are on the same maximum length. 
We use a batch size of 1 on each GPU. By sharing negative passages between GPUs, each query totally has 31 random negative passages. We train no more than 4 epochs and use well-performing learning rate for each model. Other training settings are the same as Section~\ref{sec:ms}. 

We use Transformer encoder-only, decoder-only and M2-BERT \cite{saad2024benchmarking} model as baselines. For encoder-only, we use zero-shot result of Jina Embeddings v2 model \cite{gunther2023jina} which is fine-tuned not on LoCoV0 but on a greater amount of data. Because the checkpoint of pre-trained Jina BERT is not published, we can not fine-tune it on LoCoV0. For decoder-only, we use OPT \cite{zhang2022opt} and Pythia \cite{biderman2023pythia}. For M2-BERT, we use the result in paper \cite{saad2024benchmarking} which fine-tune for 1 epoch on LoCoV0 with orthogonal projection loss \cite{ranasinghe2021orthogonal} rather than InfoNCE loss.

\begin{table}[t]
\caption{The effectiveness of Mamba Retriever on the LoCoV0 dataset compared to long-text retrievers. M2 is M2-BERT.}
\label{tab:loco}
\resizebox{\linewidth}{!}{
\begin{tabular}{l|c|c|c}
    \toprule
    \textbf{Base Model} & \textbf{Max. Len.} & \textbf{Arch.} &  \textbf{Avg. nDCG@10} \\
    \midrule
    
    M2-BERT-2k, 80M    & 2k&M2.& 83.6\\
    OPT-125M      & 2k&De.& 88.9\\
    Pythia-160M     & 2k&De.& 79.2\\
    Mamba-130M    & 2k&Ma.& \textbf{89.1}\\
    \midrule
    Jina-v2, 137M, zero-shot    & 8k&En.&85.4 \\
    M2-BERT-8k, 80M    & 8k&M2.& 85.9\\
    Mamba-130M    & 8k&Ma.& \textbf{90.7}\\
    \bottomrule
\end{tabular}
}
\end{table}

\textbf{Results.} 
Table~\ref{tab:loco} shows the effectiveness on LoCoV0. 
Briefly, Mamba shows comparable or better performance than other long-text retrievers, and can extend to longer text than pre-training. 

On 2k maximum length, although Mamba's memory capacity for long text is limited by latent state size due to the lack of self-attention mechanism, it still has comparable or better capability to Transformer decoder-only models.
In addition, although Mamba is pre-trained on 2k maximum length and M2BERT-8k is pre-trained on 8k maximum length, Mamba fine-tuned on 8k maximum length has comparable or better performance than other models. 

\subsection{Inference Efficiency}
\textbf{Implementation Details. }
We choose to measure long passage inference time which has greater potential for improvement rather than short query. We measure the inference time of all LoCoV0 train set passages \cite{saad2024benchmarking} on one A100 40G GPU using bf16, excluding tokenization time. Passages are truncated to certain lengths. We use the best batch sizes on throughput for each model and each maximum length. M2-BERT is not tested on length 512. 
Besides, generative language tasks involve multiple iterative steps, while retrieval tasks involve a single step, so there is no need to accelerate the iterative process \cite{pope2023efficiently,kwon2023efficient} in retrieval task.


\begin{figure}[t]
    \centering
    \includegraphics[width=\linewidth]{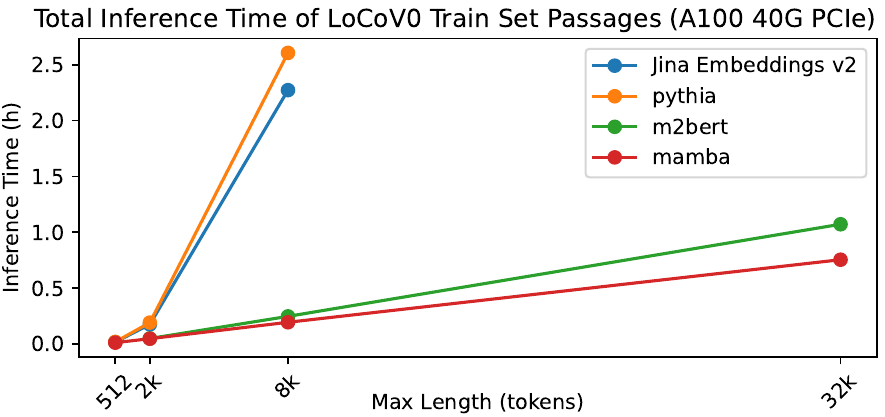}
    \caption{The efficiency of Mamba Retriever compared to long-text retrieval models at different maximum text lengths.}
    \label{fig:speed}
\end{figure}

\textbf{Results.} 
Figure~\ref{fig:speed} presents the inference time of long-text retrieval models. 
Briefly, Mamba shows faster speed on long-text retrieval. 
At maximum length of 512, the inference time of various models are similar. At 2k maximum length, Mamba and M2-BERT have similar time, while Transformer-based models require 4$\times$ time. At 8k and 32k lengths, the M2-BERT need approximately 1.2$\times$ and 1.4$\times$ inference time than Mamba, respectively.


\section{Conclusion}
In this paper, we investigate the effectiveness and efficiency of the Mamba-based model in retrieval task. The experiment results show that, on classic short-text retrieval, Mamba Retriever has similar or better effectiveness compared to Transformer retrievers, and the effectiveness increases with model size. On long-text retrieval, Mamba Retriever has similar or better effectiveness compared to existing long-text retrievers, and can extend to handle longer text lengths beyond pre-training. In addition, Mamba Retriever shows an efficiency advantage on long-text retrieval due to its linear scaling in sequence length.

\begin{acks}
This research was supported by the Natural Science Foundation of China (61902209, 62377044, U2001212), and Beijing Outstanding Young Scientist Program (NO. BJJWZYJH012019100020098), Intelligent Social Governance Platform, Major Innovation \& Planning Interdisciplinary Platform for the "Double-First Class" Initiative, Renmin University of China.
\end{acks}
\bibliographystyle{ACM-Reference-Format}
\balance
\bibliography{sample-base}


\end{document}